\def\cm2{cm$^{-2}$}

\def\c2{C~{\sc ii}}
\def\c4{C~{\sc iv}}
\def\fe2{Fe~{\sc ii}}
\def\fe3{Fe~{\sc iii}}
\def\mg1{Mg~{\sc i}}
\def\mg2{Mg~{\sc ii}}
\def\si2{Si~{\sc ii}}
\def\si4{Si~{\sc iv}}
\def\al2{Al~{\sc ii}}
\def\al3{Al~{\sc iii}}
\def\o1{O~{\sc i}}
\def\n1{N~{\sc i}}
\def\h1{H~{\sc i}}

\def\approxlt{\mathrel{\spose{\lower 3pt\hbox{$\sim$}}
        \raise 2.0pt\hbox{$<$}}}
\def\approxgt{\mathrel{\spose{\lower 3pt\hbox{$\sim$}}
        \raise 2.0pt\hbox{$>$}}}

\documentclass[article]{gwp80}
\usepackage{graphicx}
\usepackage{amssymb}
\tabletypesize{\normalsize}  

\def\plotone#1{\centering \leavevmode
\includegraphics[width=.95\columnwidth]{#1}}

\def\plotone#1{\centering \leavevmode
\includegraphics[width=.95\columnwidth]{#1}}

\shortauthors{Feast}
\shorttitle{RR Lyraes and the Magellanic Clouds}

\begin{document}
\large    
\pagenumbering{arabic}
\setcounter{page}{1}

\title{RR Lyraes and Type II Cepheids in the Magellanic Clouds: Distance Scales
and Population Gradients}
\author{{\noindent Michael Feast}\\
{\it Astronomy,Cosmology and Gravitation Centre, Astronomy Dept., University of Cape Town, 7701,
 Rondebosch, South Africa.\\
 and, South African Astronomical Observatory, P.O. Box 9, Observatory, 7935,
South Africa}}

\email{mwf@ast.uct.ac.za}

\begin{abstract}
This paper discusses the current uncertainties in luminosity calibration
of the RR Lyrae variables. The difference in distance moduli between the SMC and LMC as derived from RR Lyrae stars and classical Cepheids is used to estimate
a metallicity effect on the Cepheid PL(VI) relation  of 
$0.29\pm 0.11 \rm{mag\, dex^{-1}}$ . There is evidence that suggests RR Lyrae variables and type II Cepheids share a common $K$ - $\log P$ relation.
Metallicity and age gradients in the LMC are discussed from data on
RR Lyrae variables and AGB stars.
\end{abstract}

\section{Introduction}
The aims of the present paper are the following:\\

1. To outline briefly the current position on the luminosity calibration
of RR Lyrae variables and the future prospects.

2. To compare the relative luminosities of the RR Lyrae variables
in the LMC and SMC with that of the classical Cepheids and to deduce the
implies metallicity effect on the Cepheid scale.

3. To review infrared period-luminosity relations for
type II Cepheids and their relation RR Lyrae variables.

4. To discuss evidence for a small, but significant, mean metallicity
gradient of the RR Lyrae population in the LMC, implying a classical
picture for the formation of the LMC halo and suggesting that the
metallicity of RR Lyrae variables is correlated with their age, the
more metal-poor stars being older.

5. To suggest from published data on AGB stars in the LMC, that the oldest
stars of this type are dominant in the outer parts of the LMC whilst
the main bulk of AGB stars, which are of intermediate age, are
more centrally concentrated.   

\section{Basic Relations for RR Lyrae variables}
 In a given globular cluster RR Lyrae variables have roughly the same
$V$ magnitude independent of period though there is considerable scatter
and this increases with increasing cluster metallicity (e.g. Sandage 1990).
There is a long history of attempts to determine how $M_{V}$ depends
on metallicity. It is usually expressed in the form
\begin{equation}
M_{V} = \alpha [Fe/H] + \beta
\end{equation}
but it is not clear whether it is linear over the full range of possible
values of [Fe/H] (see e.g. Feast 1999, McNamara 1999).

Probably the best determination of the slope of this relation is from
the work of Gratton et al. (2004). They obtained,
\begin{equation}
V_{o} = 0.214 (\pm 0.05)([Fe/H] +1.5) + 19.064
\end{equation}
from LMC data with the values of [Fe/H] being determined by a modification
of the Preston (1959) method.  Whilst this slope agrees well with that
found, for instance, in our Galaxy using pulsation parallaxes, a smaller slope
($0.09 \pm 0.03$) was found in the Sculptor dwarf spheroidal (Clementini et al. 2005). These authors suggested that this was due to the Sculptor variables
being on average more evolved than those in the LMC. Evolution may also be part of the reason for the spread in these relations (a total spread of
$\sim 0.5$mag in the case of the LMC) though a significant amount is likely
to be due to the depth of the LMC and Sculptor.

Longmore et al. (1986) found that RR Lyraes in globular clusters followed 
a $K$ (2.2 microns) versus $\log$ period relation. The scatter in this 
relation at a given metallicity is small. For instance in the case of the Reticulum cluster in 
the LMC the standard deviation about such a relation is only 0.03mag (Dall'Ora et al. 2004).
The relation may be written,
\begin{equation}
M_{K} =\gamma \log P + \delta [Fe/H] + \phi,
\end{equation} 
where a term has been included for a possible metallicity dependence.
Table 1 contains a number of recent estimates of $\gamma$.
\begin{flushleft}
\begin{deluxetable*}{ll}
\tabletypesize{\normalsize}
\tablecaption{The slope $\gamma$ in the RR Lyrae $M_{K} -\log P$ relation}
\tablewidth{0pt}
\tablehead{\\ \colhead{Source} & \colhead{$\gamma$}}\\
\startdata
Globular clusters (Sollima et al. 2006) & $-2.38 \pm 0.04$\\
Reticulum cluster (Dall'Ora et al. 2004) & $-2.16 \pm 0.09$\\
LMC Field (Borissova et al. 2009) & $-2.11 \pm 0.17$ \\
LMC Field (Szewczyk et al. 2008)  & $-2.19 \pm 0.40$\\
SMC Field (Szewczyk et al. 2009) & $-3.10 \pm 0.49$\\
Theory (Bono et al. 2003) & $-2.10$\\
Theory (Catelan et al. 2004) & $-2.35$ \\
\enddata
\end{deluxetable*}
\end{flushleft}
The result from Sollima et al.(2006) is the mean from a number of Galactic
globular clusters. 
The scatter about a mean relation is  much larger in the LMC and SMC fields than in individual
globular clusters
(Szewczyk et al. 2008, 2009). This is probably mainly due to the depth of these
galaxies but the range in [Fe/H] may also contribute as well as the lack
of full $K$ light curves. 

An estimate of the coefficient, $\delta$, of the metal term in eq. 3 was made by Sollima et al. (2006) using globular clusters of different metallicities
with distances derived from main sequence fitting. They found
$\delta = +0.08 \pm 0.11$.  A similar value, $+0.05\pm 0.17$, 
was obtained by Borissova et al.
(2009) from RR Lyraes with known values of [Fe/H] in the LMC.
These two values are not significantly different from 
zero but agrees within the errors with the theoretical estimates of
Bono et al. (2003) (+0.23) and of Catelan et al. (2004) (+0.18).
\section{The SMC-LMC Modulus difference from RR Lyraes and classical Cepheids}
Whilst the absolute luminosity scale for classical Cepheids of close to solar metallicity has been fixed, at least at the shorter periods, by
trigonometrical parallaxes (Benedict et al. (2007); van Leeuwen et al. (2007)),
there is still considerable uncertainty in the effects of metallicity
on the scale. 
Matsunaga et al. (2011) have compiled data
on the relative distances  of the SMC and LMC and this is relevant to the
metallicity issue. Table 2 is based on their work.

\begin{flushleft}
\begin {deluxetable*}{lll}
\tabletypesize{\normalsize}
\tablecaption{SMC-LMC Modulus Difference}
\tablewidth{0pt}
\tablehead{\\ & \colhead{RR Lyraes ($K$)} & \colhead{Cepheids ($VI$)}}\\
\startdata
Uncorr. & $0.327 \pm 0.002$ & $0.48 \pm 0.01$\\
Corr. & $0.363 \pm 0.04$ & \\

& &\\
$\Delta$[Fe/H] & $-0.22$ & $-0.42 \pm 0.15$\\
\enddata
\end{deluxetable*}
\end{flushleft}
In the case of the RR Lyrae variables the data are from Szewczyk et al. (2008,
2009) based on the PL(K) relations 
with the difference in mean metallicities which they adopt. 
The table lists the modulus difference without metallicity correction and the metallicity corrected value, using a mean of the various estimates
of $\delta$ discussed in the last section.
 Evidently the metallicity correction has rather little effect on the modulus difference unless the metallicity difference between the Clouds and/or
the coefficient, $\delta$, in eq. 3, have been grossly
underestimated.
From these data we adopt a modulus difference of 0.36 
for the Clouds. 
 
The results in table 2 for the (classical) Cepheids are for the PL(VI) 
relations uncorrected for
metallicity effects and are from the Appendix of Matsunaga et al. (2011).
The Cepheid metallicity difference quoted is based on the spectroscopic determination of iron abundances in both Clouds by Romaniello et al. (2008).
To bring the Cepheid modulus difference to the value given by the 
RR Lyrae variables requires a metallicity correction equivalent to
$0.29\pm 0.11 \rm{mag\,dex^{-1}}$.
This happens to be in exact agreement with the the value  derived
by Macri et al. (2006) from observations of Cepheids in NGC4258,
$0.29 \pm 0.10 \rm {mag\,dex^{-1}}$. 
The metallicity
correction taken from Macri et al. is derived using metallicities 
of HII regions measured on the empirical scale of Zaritsky et al (1994).
On the $\rm{T_{e}}$ scale of Kennecutt et al (2003), which has been extensively
used in extragalactic work, the correction would be greater 
($0.49\pm 0.15 \rm{mag\,dex^{-1}}$) and the corrected Cepheid SMC-LMC modulus difference would be 0.30mag, agreeing less well with the RR Lyraes.
Bono et al. (2010)  obtained a Cepheid metallicity effect on PL(VI) of 
$0.03\pm 0.07 \rm{mag\,dex^{-1}}$ on the basis of galaxy distances
derived from tip of the RGB magnitudes. This agrees with their theoretical
estimate that the metallicity effect in PL(VI) is small.

A caveat in the discussion of the SMC-LMC difference, is that it depends on the assumption
that the distribution of RR Lyraes and Cepheids in each Cloud is such
that their mean distances are the same. Provided the metallicity effect
of $0.29\pm 0.11 \rm{mag\,dex^{-1}}$ can be extrapolated linearly to higher metallicities, the
correction to the LMC modulus based on Cepheids of near solar metallicity
is $-0.09 \pm 0.05$ mag. Adopting an uncorrected Cepheid modulus of
$18.52\pm 0.03$ from van Leeuwen et al. (2007), the corrected modulus is
$18.43\pm 0.06$. Evidently the metallicity correction remains the most significant uncertainty in the Cepheid distance to the LMC.

\section{RR Lyrae Absolute Magnitudes}
This section discusses the values of $\beta$ and $\phi$ in the
equations,
\begin{equation}
M_{V} = 0.21 [Fe/H] +\beta
\end{equation}
and
\begin{equation}
M_{K} = -2.33 \log P + \phi.
\end{equation}
The value of the coefficient of [Fe/H] is from Gratton et al (2004) (see above) and the coefficient of $\log P$ is one that has been used by various workers and is consistent with the discussion of section 2. There are
three methods which have been used to establish  RR Lyraes as primary distance
indicators; trigonometrical parallaxes, statistical parallaxes and 
pulsation parallaxes. The available data is summarized in Table 3.
The results from trigonometrical parallaxes rely entirely on the the HST
parallax of RR Lyrae itself (Benedict et al. 2002,
Feast et al. 2008)). Other parallaxes for
this type of star are too poor to add significantly to the result . 
The most elaborate study of RR Lyrae statistical parallaxes is
that of Popowski and Gould (1998a, b, c). Their results
lead to the value of $\beta$ in the table. The value of $\phi$ from
statistical parallaxes is from Dambis (2009).  There have been a considerable
number of determinations of absolute magnitudes of RR Lyrae variables from
pulsation parallaxes. The results depend on the models adopted (see for
instance the discussion by Cacciari \& Clementini (2003).  The pulsation
parallax results of Fernley et al. (1998) lead to the tabulated value of
$\beta$ whilst the value of $\phi$ is derived from the data of
Jones et al. (1992).

A striking feature of Table 3  is that the values of $\beta$ and $\phi$
derived from statistical and pulsation parallaxes agree closely, whilst
the trigonometrical parallax result of RR Lyrae itself is discrepant.
This is particularly notable in the case of $\phi$.  The position is clearly unsatisfactory. Fortunately the preliminary
results of the new HST trigonometrical parallax programme (Barnes, this
volume) indicate that the calibration will soon be greatly improved. 
Any remaining difference between the trigonometric and the statistical
result would be of considerable interest as it might indicate that the Galactic model use in the statistical work was unsatisfactory.
It is clear, for instance from the work of Martin \& Morrison (1998),
that the mean velocity of halo RR Lyrae variables relative to the Sun in the
direction of Galactic rotation is quite sensitive to the absolute magnitudes
adopted. A difference between accurate trigonometrical parallaxes and
pulsation parallaxes would indicate a need to update RR Lyrae models.
\begin{flushleft}
\begin{deluxetable*}{lll}
\tabletypesize{\normalsize}
\tablecaption{ RR Lyrae zero-points}
\tablewidth{0pt}
\tablehead{\\ \colhead{Method} & \colhead{$\beta$} &  \colhead{$\phi$}}\\
\startdata
Trig. Par. & $+0.52 \pm (0.11)$ & $-1.22 \pm (0.11)$\\
Stat. Par. & $+0.79 \pm 0.13$ & $-0.82 \pm 0.08$\\
Puls. Par. & $+0.73 \pm 0.14$ & $-0.88 \pm 0.06$\\

\enddata
\end{deluxetable*}
\end{flushleft}

\section{Type II Cepheids}
\begin{figure*}
\centering 
\plotone{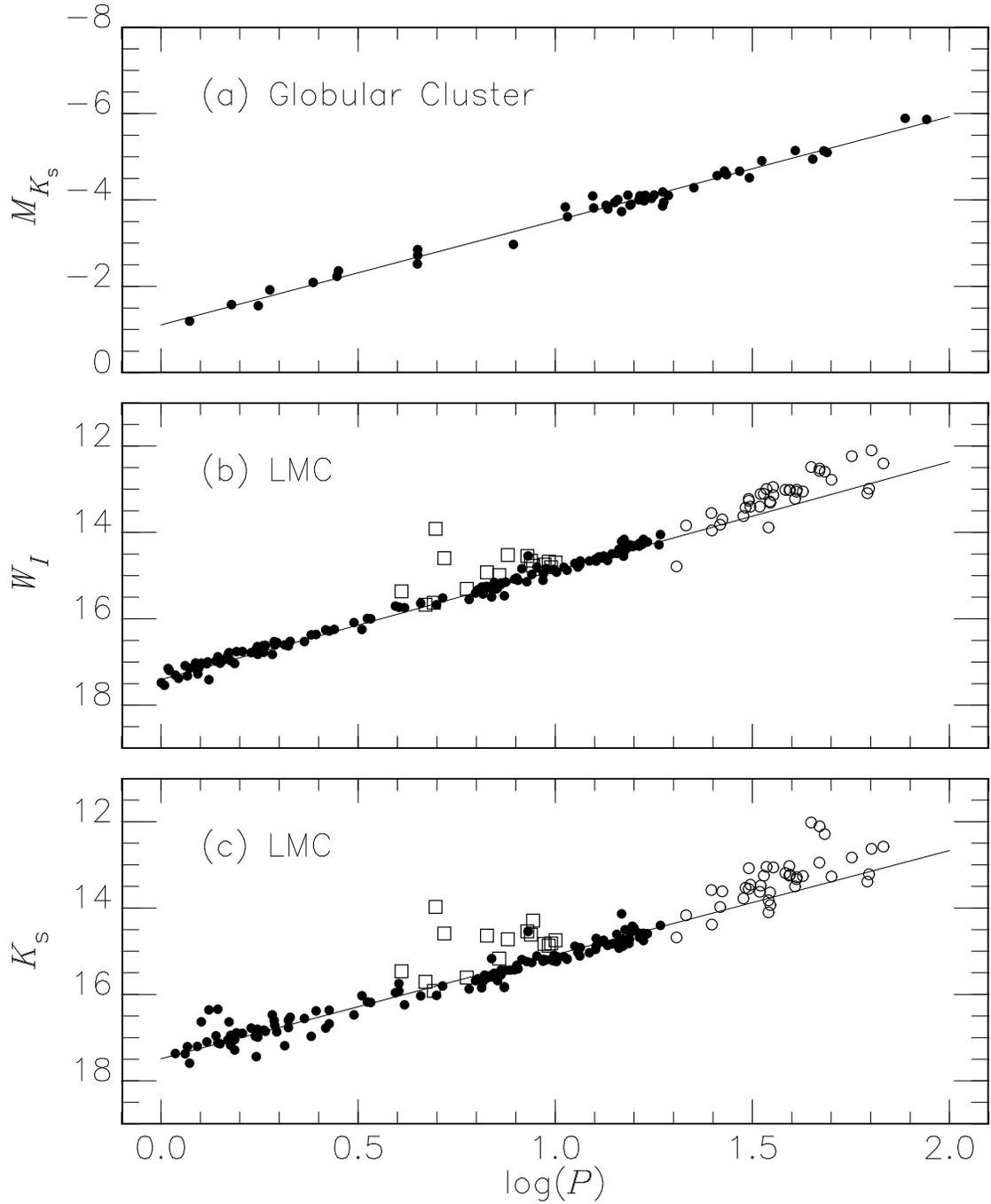}
\caption{Period-luminosity relations for type II Cepheids. In b and c
filled and open circles are for objects with periods below and above
20 days. The open squares are for the peculiar W Vir stars. See
text for discussion}
\end{figure*}

Like the RR Lyrae variables, the type II Cepheids belong to both
halo and old disc populations.
Matsunaga et al. (2006) showed that the type II Cepheids in globular
clusters follow well defined period -luminosity relations in
$J,H$ and $K$. This is illustrated in fig 1. The figure also shows the
period-luminosity relations for this type of variable in the LMC
at W(VI) as obtained by Soszy\'{n}ski et al. (2008) and at $K_{s}$ 
by Matsunaga et al. (2009) for the
same stars using the IRSF point source catalogue of the Magellanic Clouds
(Kato et al. 2007). Similar results have been obtained for the SMC
(Matsunaga et al. 2011). 
The main difference between the globular cluster
results and those for the LMC and SMC is the presence of some stars in the
W Vir period range (periods greater than 4 days and less than 20 days) 
above the period-luminosity relations in the Clouds. Soszy\'{n}ski et al. (2008)
show that such stars have distinctive light curves. 
Also, at the long period
end (the RV Tau period range, periods greater than 20 days) most of the stars in the LMC and SMC lie above
the period-luminosity relations. Further work is required to
see whether there are stars in this period range in the LMC and SMC
which are similar to those in globular clusters. In addition there is some
suggestion that the slope of the period-luminosity relation  at K (omitting
stars in the RV Tau range) may vary from system to system (Matsunaga et al.
2011)(see table 4). Further work on this is required. It might, for instance, indicate a period dependent metallicity effect. There might also
be problems with selection effects at the short period end.

The slopes for the type II Cepheid period-luminosity relation in table 4
are very similar to those give for the RR Lyraes in table 1. Furthermore,
Matsunaga et al. (2006) showed that within the uncertainties of relative
distance estimation, the RR Lyraes in the globular cluster NGC6341 fitted
an extrapolation of the globular cluster type II Cepheid $K$-band period-luminosity to shorter periods. It is therefore possible that there is a common period-luminosity relation covering both types of variable.

\begin{flushleft}
\begin{deluxetable*}{ll}
\tabletypesize{\normalsize}
\tablecaption{Slopes of the type II Cepheid $M_{K} - \log P$ relation}
\tablewidth{0pt}
\tablehead{\\ \colhead{System} & \colhead{Slope}}\\
\startdata
Globular Clusters & $-2.41 \pm 0.05$\\
LMC Field & $-2.28 \pm 0.05$\\
SMC Field & $-2.11 \pm 0.10$\\
\enddata
\end{deluxetable*}
\end{flushleft}
 
\section{A Metallicity gradient in the LMC RR Lyrae population}
\begin{figure*}
\centering
\plotone{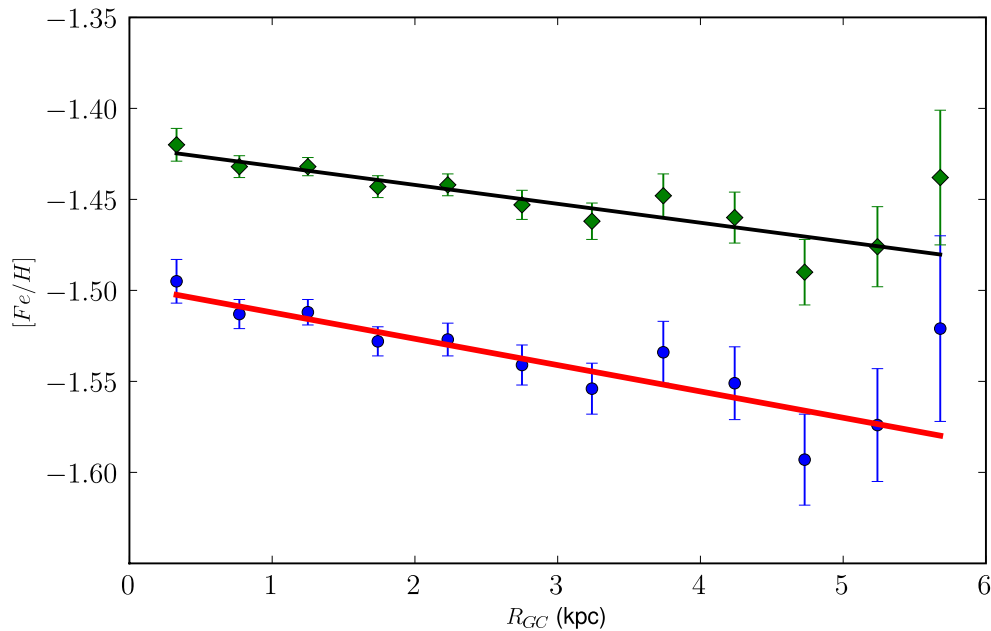}
\caption{The relation between [Fe/H] and $\rm R_{GC}$, the distance from the
centre of the LMC for RR Lyrae variables. The two lines are for two
slightly different relations between [Fe/H] and period}
\end{figure*}
Our understanding of the formation and evolution of dwarf galaxies such as
the LMC is rather sparse. The RR Lyraes, as representing the oldest
populations, are particularly important in this regard. This field has
been revolutionized by the work of the OGLE group. The OGLE-III catalogue
(Soszy\'{n}ski et al. (2009) lists 17,693 variables of type RRab. Leaving out
those that are likely to be blended or foreground or are otherwise dubious,
there is a sample of 16,864 RRab stars available for analysis. Feast et al.
(2010) have used these data to study the change of mean period with distance
($\rm R_{GC}$) from the centre of the LMC. Fig. 2 shows the results, with mean period converted to mean metallicity using two possible (Galactic) mean period -metallicity relations. The gradient is small but significant. For instance
the upper line in fig. 2 has the equation,
\begin{equation}
[Fe/H] = -0.0104(\pm 0.0021)R_{GC} - 1.4213 (\pm 0.0046)
\end{equation}
It should be noted that fig. 2 simply shows  linear, scaled versions, of
a mean period  versus $R_{GC}$ relation and this relation remains if one
chose to discount the mean relation between period and metallicity.
The period gradient, though slight, indicates that the oldest
populations in the LMC have a clear structure 
with the most metal poor component showing the greatest extent.
This would , for instance,
be consistent with the classical picture of formation by the collapse
of a gas cloud.

\vfill\eject
\section{An Age Gradient in the AGB star population of the LMC}
\begin{figure*}
\centering
\plotone{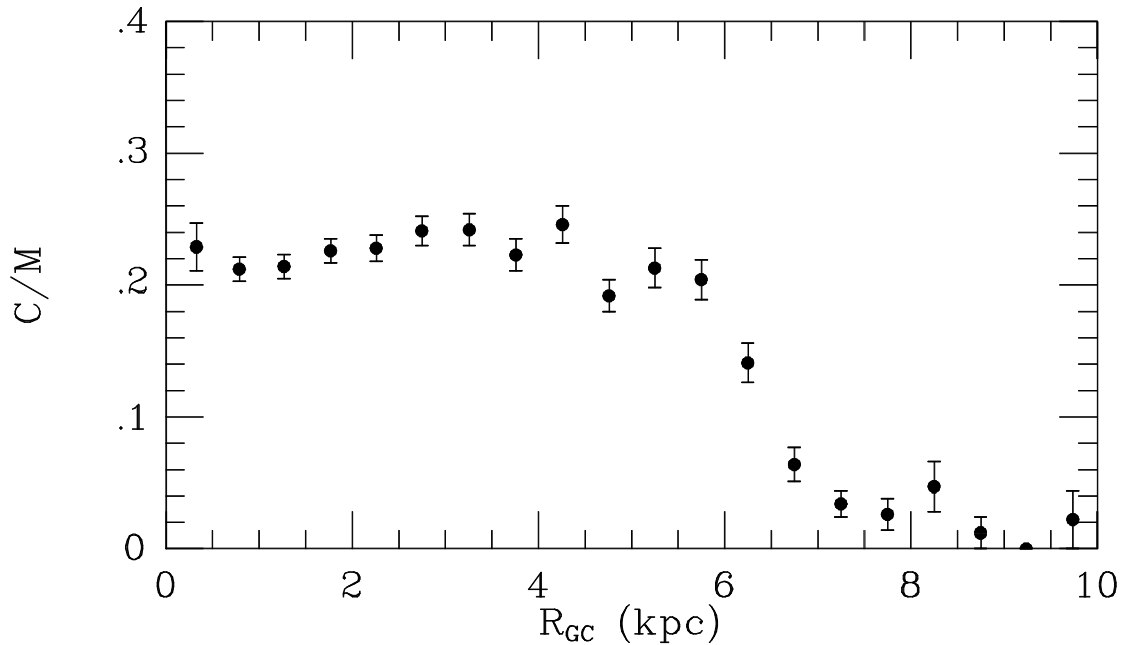}
\caption{The ratio of the number of photometrically selected  carbon -rich AGB stars to those of type M as a function of distance from the centre of
the LMC}
\end{figure*}
It is of interest to ask if there are metallicity or
age gradients in the LMC for populations other than the RR Lyraes. In the case of the youngest populations,
HII regions give only very marginal evidence for a metallicity gradient
(Pagel et al. 1978). AGB stars belong to intermediate and old populations.
Cioni \& Habing (2003) have selected LMC AGB stars (i.e. they a brighter
than an adopted RGB tip) and have divided then into probable C and M type
stars using DENIS $IJK$ colours. Dr Cioni kindly made the data (32,801 stars)
available and in fig 3. the number ratio of C/M stars is plotted
as a function the distance from the LMC centre (Feast et al. 2010).
There is some very slight evidence of a small increase in C/M out to
$\sim 4$kpc, beyond which there is a steep drop. In the past the C/M ratio has been 
thought to increase with decreasing metallicity and this may indeed be the case when comparing systems of about the same age. However it seems quite
unlikely that there is a marked increase in metallicity as one 
moves outwards beyond 
4kpc. It seems much more likely that the outer AGB population is dominated
by older stars. It should be noted that this decrease in C/M ratio is not
affected by the possible inclusion of galactic foreground stars. The density of the selected AGB stars at $R_{GC} \sim 6$kpc, whilst $\sim 20$
times lower than in the centre is still $\sim 25$ times greater than at
10kpc where it is still falling.

\section{Conclusions}
At present the absolute calibration of RR Lyrae luminosities is in a
very unsatisfactory state since the results from statistical parallaxes
differ by several tenths of a magnitude from that implied by the trigonometrical parallax of RR Lyrae itself. This matter should be clarified
soon by the current parallax work of Benedict et al.

The difference in distance moduli between the SMC and LMC as derived from
RR Lyraes and classical Cepheids suggests that the Cepheid scale is
metallicity dependent unless there is a difference in the  mean distance of the two types of variables in the SMC and/or the LMC.

Type II Cepheids show period-luminosity relations in the infrared and in
$VI$. The extension
of the type II PL(K) relation to shorter periods may well fit the RR Lyrae variables.

There is evidence of a small metallicity gradient in the RR Lyrae population
of the LMC consistent with the the classical picture of LMC formation
by collapse of a gas cloud. An age gradient is present in the AGB star
population
\section{Acknowledgements}
 I am grateful to Dr Noriyuki Matsunaga for providing figure 1 and
to him and Prof. Patricia Whitelock for
discussions on the topics of this paper.

\end{document}